\documentstyle[epsf]{l-aa}
\newcommand{\cyg}{\mbox{Cyg~X--1}}
\newcommand{\rms}{\mbox{r.m.s.}}

\begin{document}
\thesaurus{06         
           (13.25.5;  
            08.02.3;  
            02.02.1;  
            02.01.2)} 

\title{1100 Days of BATSE Observations of Cygnus X-1}

\author{D.J. Crary\inst{1}
\and    C. Kouveliotou\inst{2}
\and    J. van Paradijs\inst{3,4}
\and    F. van der Hooft\inst{4}
\and    D.M. Scott\inst{2}
\and    S.N. Zhang\inst{2}
\and    B.C. Rubin\inst{2}
\and    M.H. Finger\inst{2}
\and    B.A. Harmon\inst{5}
\and    M. van der Klis\inst{4}
\and    W.H.G. Lewin\inst{6}}

\institute{NAS/NRC Research Associate, NASA Code ES-84, Marshall Space Flight Center, Huntsville, AL\ \ 35812, USA
\and Universities Space Research Association, Huntsville, AL\ \ 35806, USA
\and Department of Physics, University of Alabama in Huntsville, Huntsville, AL\ \ 35899, USA
\and Astronomical Institute ``Anton Pannekoek'', University of Amsterdam \& Center for High-Energy Astrophysics, Kruislaan 403, NL-1098 SJ Amsterdam, The Netherlands
\and NASA/Marshall Space Flight Center, Huntsville, AL 35812, USA
\and Massachusetts Institute of Technology, 37-627 Cambridge, MA 02139, USA}

\offprints{D.J. Crary}

\date{Received date; accepted date}

\maketitle

\begin{abstract}
We have examined the power spectral behavior of \cyg\ using approximately
1100 days of BATSE data.  These data have been searched for correlations
between different power spectral features, and for correlations between 
power spectral features and energy spectral parameters derived from BATSE 
occultation analysis.
\end{abstract}

\section{Introduction}

A data base has been created combining 1100 days of \cyg\ observations
with BATSE, from JD 2\,448\,387 to 2\,449\,480 
(\mbox{May 11, 1991} to \mbox{May 8, 1994}).
It includes observations of fast variability (0.01--0.488 Hz) as well as
flux measurements (\mbox{45--140 keV}) and energy spectral fits. The goal of
this work is to look for correlations between various parameters derived
from power spectral, flux, and energy spectral measurements.  This work 
is part of a larger on-going project to systematically study the
power spectral behavior of all black-hole candidate sources observed 
with BATSE.

In Section 2 we describe the methods used for creating the power spectra
and the results of the occultation analysis.  In Section 3 we show
that correlations occur between several quantities derived from the 
energy spectral and power spectral fits, and briefly discuss these results.

\section{Observations}

The BATSE occultation analysis technique has been described elsewhere
(Harmon et al.\ \cite{harm:occ}).  For this study, the corrected detector count
rates for \cyg\ have been fit to a power law in the energy range 45--140 keV,
and a total flux (photons $\rm cm^{-2}\: sec^{-1}$) has been 
calculated for the same
energy band.
Typically 15--30 measurements are made each day.
From these
data we produced daily averaged flux and photon spectral \mbox{index values.}

To quantify the rapid variability, we have created power spectra from
the 1.024 second time resolution large area detector (LAD) count rate
data, using two energy channels covering the range 20--50 and 50--100 keV.
These data were filtered to eliminate bursts, then
searched for segments with 512 contiguous time bins (524.288 seconds without
gaps) when the source was above the Earth's limb.  Each segment was
fit to a quadratic polynomial and the fit residuals converted to
a power density spectrum (PDS) using standard fast Fourier 
transform techniques.
The PDS was normalized to squared fractional \rms\ amplitude
per unit frequency, according to the method described by 
Miyamoto et al.\ (\cite{miya:mus}), using the raw daily averaged 
detector count rates in the 20--100 keV energy band obtained
from the occultation analysis.  Interference from
other black hole candidate sources (e.g., \mbox{GRO~J1719$-$24}, 
\mbox{GRO~J0422+32} and \mbox{GX 339$-$4}) was
eliminated by selection of only those detectors
in which these sources did not appear.

Using similar 524.288 second intervals obtained when the source was occulted
by the Earth, we have determined that the quadratic detrending of the raw
data yields a background (source occulted) power level consistent with a
pure Poisson process for frequencies between 0.01 Hz and the Nyquist frequency
(0.488 Hz).  The results obtained are consistent with those obtained using
a full background model developed by Rubin et al.\ (\cite{rub:bck}).
\begin{figure}
\epsfysize=6.5cm
\epsffile{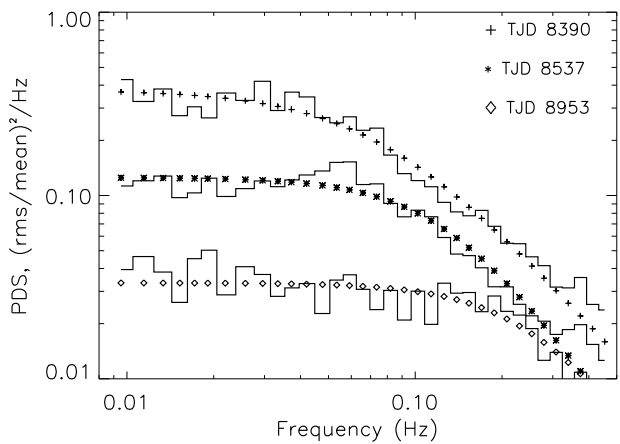}
\caption{Randomly selected daily averaged PDS and fits to the functional
form described in the text; the data have been rebinned into approximately
equal logarithmic intervals} 
\end{figure}
\begin{figure}
\epsfysize=6.5cm
\epsffile{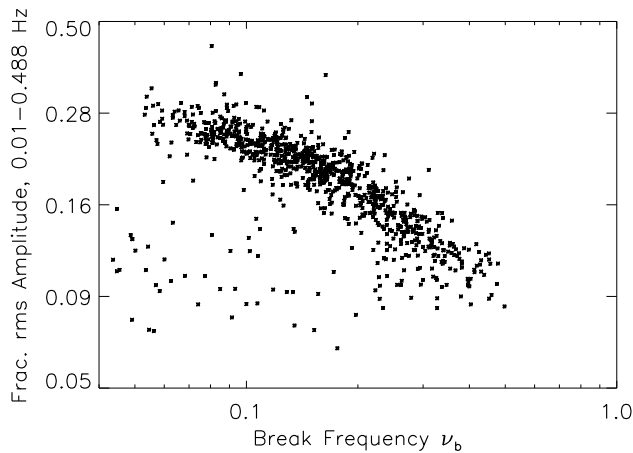}
\caption{Observed correlation between the break frequency, $\rm \nu_b$, and
\rms\ amplitude}
\end{figure}

It is well known that the power spectrum of \cyg\  has a characteristic 
shape (Belloni \& Hasinger \cite{bell:has}); the spectrum is flat below
a \lq break' frequency, $\rm \nu_b\sim\! 0.1\, Hz$,  and has 
a power law form above
$\rm \nu_b$.  We have attempted to parameterize the shape of these spectra
by fitting to a function, P, of the \pagebreak form
\begin{equation}
\lefteqn{{\rm P}(\nu)=\frac{a_0}{1+(\frac{\nu}{a_1})^{a_3}}.} \label{eq:pow}
\end{equation}
Here $a_0$ is the amplitude of the low frequency part of the spectrum, 
$a_1\!\simeq\!\nu_{\rm b}$, and $a_3$ is related to the slope of the 
power spectrum just above the break frequency.
Figure 1 shows three randomly selected spectra 
and the fits obtained using this functional form.  We also calculated the
fractional \rms\ amplitude from the power
spectral density for the frequency range \mbox{0.01--0.488 Hz}.

\section{Results}

The three power spectral fit parameters ($a_0$, $a_1$, $a_2$), 
the fractional \rms\ amplitude, the the total flux, 
and photon power-law index were plotted against one 
another to study the relationships between these 
quantities. Some of the results are shown in Figs.~2 and 3.

\begin{figure}
\epsfysize=6.5cm
\epsffile{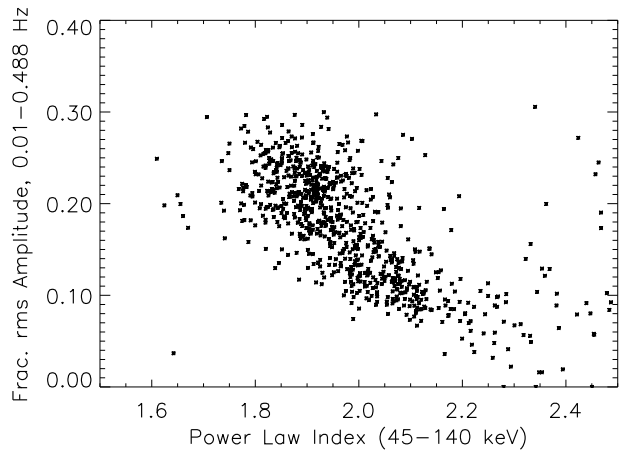}
\caption{Observed correlation between the spectral index of a power-law
fit to the energy spectrum between 45--140 keV, and the \rms\ amplitude}
\end{figure}
 
The strongest correlation
is observed between the break frequency $\rm \nu_b$ ($a_1$) and the
\rms\ amplitude (Fig. 2). 
The fractional \rms\ amplitude monotonically decreases as the 
$\rm \nu_b$ increases.
These observations are similar to the results of
Belloni \& Hasinger (\cite{bell:has}) based on 13 observations
from the EXOSAT archive at lower energies ($\sim\!$ 2--20 keV).

Figure 3 shows a plot of fractional \rms\ amplitude vs. spectral
index of a power law fit in the 45--140 keV band.  A correlation
is also seen in these data.  The energy spectrum generally becomes softer 
as the
\rms\ amplitude decreases.  The fractional \rms\ amplitude is in the
range 8--30\% during these observations. The origin of this effect is 
not well understood.  A more detailed
study of the type presented here is in preparation.

\begin{acknowledgements} 
This project was performed within NASA grant NAG5-2560 and supported
in part by the Netherlands Organization for Scientific Research (NWO)
under grant \mbox{PGS 78-277.}
FvdH acknowledges support by the Netherlands Foundation for Research in
Astronomy with financial aid from NWO under contract number
\mbox{782-376-011}. JvP acknowledges support from NASA grant NAG5-2755.
\end{acknowledgements}

\end{document}